# 4927 | Tip angle dependence for resistive force into dry granular materials at shallow cone penetration

Naoki Iikawa [a, b, c, *] and Hiroaki Katsuragi [c]

[a] *Development Division, Komatsu Ltd, 3-25-1 Shinomiya, Hiratsuka, 254-0014, Kanagawa, Japan*

[b] *Komatsu MIRAI Construction Equipment Cooperative Research Center, Osaka University, 1-1 Yamadaoka, Suita, 565-0871, Osaka, Japan*

[c] *Department of Earth and Space Science, Osaka University, 1-1 Machikaneyama, Toyonaka, 560-0043, Osaka, Japan*

\* *Corresponding author: naoki_iikawa@global.komatsu*

**A B S T R A C T**

In relation to the interaction of the earth's surface with machines and organisms, and its engineering applications, there has been a recent increase in interest in the penetration resistive force into granular materials at shallow depths. Previous studies have proposed various models for penetration resistive forces into dry granular materials. This study focuses on the model which has a coefficient depending on the angle of repose and the increase of resistive force in proportion to the penetration volume. In the previous studies, the model has been validated for several geometries such as cylinders, cones, and spheres. However, for cones, the model has only been validated under conditions of a tip angle close to the angle of repose. In this study, the effect of cone tip angle on penetration resistive force is investigated under several conditions with different angles of repose. This study carries out cone penetration simulations using the discrete element method. For the cone geometry, five tip angles ranging from sharp to blunt (tip angles are 15, 30, 45, 60, 75 deg) are used. The simulation results show that the penetration resistive forces for blunt cones are much higher than that computed by the model. To solve the discrepancy between the model and simulation results, this study modifies the model by assuming that the stagnant zone formed in front of the cone penetrating the granular material behaves as an effective cone. Thereby, the proposed model can calculate penetration resistive forces more accurately for cones with a wider range of tip angles than in the previous model.

*Keywords*
Soil characterization
Granular materials,
Discrete-element modelling,

## 1. Introduction

Recent planetary explorations have revealed that the surfaces of the Earth and several other solid planets are covered with granular materials. To understand and predict the motions of machines and organisms on these surfaces (e.g. landing (Ballouz et al. 2021), locomotion (Suzuki et al. 2023), and excavation (Miyai et al. 2019)), the interaction between objects and granular materials needs to be clarified. These phenomena vary in behavior but are in common that they include object penetration into granular materials. Thereby, it is important to estimate the penetration resistive force during object penetration into granular materials.

Previous studies have proposed various models to estimate penetration resistive force. In these models, the phenomenological models have been often used to understand the impact crater dynamics and mechanical properties of regolith in Planetary Science (Katsuragi, 2016; Okubo & Katsuragi, 2022), and to comprehend the ecology and morphology of organisms on sand in Biology (Sharpe et al., 2015). These models are generally expressed as the sum of a depth-dependent term derived from hydrostatic-like pressure and a velocity-dependent term derived from viscosity and/or inertia (Katsuragi, 2016). In Terramechanics, Resistive Force Theory (RFT) which calculates the penetration resistive force for each infinitesimal surface of the intruder on each orientation relative to the ground and direction of movement, is often used (Li et al., 2013). In fact, recent study has performed simulations which reproduce the machine behaviors on granular materials by combining with RFT and Multi-Body Dynamics (Suzuki et al. 2023). While these models can accurately and quickly calculate the resistive force, it is necessary to calibrate coefficients in advance. Therefore, it is difficult to predict resistive forces when the ground material cannot be used previously.

Among these studies, a new model based on the classical theory in Geotechnical Engineering has been proposed recently (Kang et al., 2018). This model is called as modified Archimedes' law theory (MALT) and described as follows:

$$\boldsymbol{F_z}(z_p) = K_\phi \rho_g \psi \boldsymbol{g} V_p(z_p),$$
$$\text{with } K_\phi \equiv \left(2 \frac{1+\sin\phi}{1-\sin\phi} e^{\pi \tan\phi} \int_0^1 \eta A(\eta, \phi) d\eta\right), \quad (1)$$

where $\boldsymbol{F_z}$ is the resistive force perpendicular to the layer; $z_p$ is the penetration depth; $\rho_g$ is the density of granular particles; $\psi$ is the packing fraction (which equals 1 minus the void ratio); $\boldsymbol{g}$ is the gravitational acceleration; $V_p$ is the penetration volume of the intruder; $K_\phi$ is the coefficient which depends only on the angle of repose $\phi$. The details of parameters $\eta$ and $A(\eta, \phi)$ in *Eq.* (1) are explained in Kang et al. (2018). MALT can estimate $\boldsymbol{F_z}$ without prior calibration (it only requires the physical properties of granular layers). However, MALT has an issue to apply them to the estimation of resistive force in actual soil ground. It is the effect of the intruder shape. Kang et al. (2018) have validated MALT with respect to several intruder geometries such as cylinders, spheres, and cones, but only for cone tip angles close to $\phi$. On the other hand, Mishra et al. (2018) have reported that $\boldsymbol{F_z}$ decreases as the cone tip shape becomes sharper. Thus, it is not clear whether MALT is applicable for various cone tip shapes.

In this study, the effect of cone tip angle on penetration resistive force in dry granular layers is investigated through Discrete Element Method (DEM) simulations. Using the simulation results, this study discusses the validity of MALT for various cone tip angles and proposes the modified model which can calculate penetration resistive forces more accurately for cones with a wider range of tip angles.

## 2. Materials and Methods

### 2.1. Discrete Element Method (DEM)

This study uses an open-source DEM engine, LIGGGHTS(R)-PUBLIC Version 3.8.0 (Kloss and Goniva 2011; Berger et al. 2015) to perform the DEM simulations. In the simulations, the equations of motion for the translational and rotational directions of the particle $i$ are respectively expressed by the following equations:

$$m_i \frac{d\mathbf{v}_i}{dt} = \sum (\mathbf{F}_n^j + \mathbf{F}_t^j) + m_i \mathbf{g},$$
$$I_i \frac{d\boldsymbol{\omega}_i}{dt} = \sum (r_i \times \mathbf{F}_t^j + \mathbf{M}_r^j + \mathbf{M}_s^j), \quad (2)$$

where $m_i$, $\mathbf{v}_i$, $r_i$, $I_i$, and $\boldsymbol{\omega}_i$ are mass, velocity, radius, moment of inertia, and angular velocity of particle $i$, respectively; $\mathbf{F}_n^j$ and $\mathbf{F}_t^j$ are respectively normal and tangential forces between particle $i$ and $j$; $\mathbf{M}_r^j$ and $\mathbf{M}_s^j$ are respectively rolling and twisting moments from particle $j$ to $i$; the symbol $\Sigma$ denotes the sum of all forces or moments acting on particle $i$ from adjacent particles.

In Eq. (2), $\mathbf{F}_n^j$ and $\mathbf{F}_t^j$ fully follow the Hertz model in LIGGGHTS. Each force is described as follows:

$$\mathbf{F}_n^j = k_n \boldsymbol{\delta}_{nij} + \gamma_n \mathbf{v}_{nij},$$
$$\mathbf{F}_t^j = \min[\mu |F_n^j|, k_t \boldsymbol{\delta}_{tij} + \gamma_t \mathbf{v}_{tij}], \quad (3)$$

where $\boldsymbol{\delta}_{nij}$ and $\boldsymbol{\delta}_{tij}$ are respectively normal and tangential overlap displacements between particle $i$ and $j$; $\mathbf{v}_{nij}$ and $\mathbf{v}_{tij}$ are respectively normal and tangential components of the relative velocity between particle $i$ and $j$; $\mu$ represents the sliding friction coefficient. Here, $\mu_{gg}$ and $\mu_{og}$ are respectively used in the case for particle-particle contact and object-particle contact. The symbol "min[ ]" denotes choice of the smaller of the two values in parentheses. The details of the elastic constants for normal contact $k_n$ and tangential contact $k_t$, and the viscoelastic damping constants for normal contact $\gamma_n$ and tangential contact $\gamma_t$ are explained in the documentation of LIGGGHTS(R)-PUBLIC v3X (https://www.cfdem.com/media/DEM/docu/gran_model_hertz.html, the last accessed: April 22, 2024).

In addition, $\mathbf{M}_r^j$ and $\mathbf{M}_s^j$ in Eq. (2) also fully follow the rolling resistance model proposed by Jiang et al. (2015). Each moment is described as follows:

$$\mathbf{M}_r^j = \min[0.525|F_n^j|\chi r^*, 0.25(\chi r^*)^2(k_n \boldsymbol{\delta}_{\theta ri} + \gamma_n \boldsymbol{\omega}_{ri})],$$
$$\mathbf{M}_s^j = \min[0.65\mu |F_n^j|\chi r^*, 0.5(\chi r^*)^2(k_t \boldsymbol{\delta}_{\theta si} + \gamma_t \boldsymbol{\omega}_{si})], \quad (4)$$

where $\chi$ is numerical parameter expressing rotation resistance due to particle surface irregularities; $\boldsymbol{\delta}_{\theta ri}$ and $\boldsymbol{\delta}_{\theta si}$ are respectively total relative rolling and twisting angular displacements from particle $j$ to $i$; $\boldsymbol{\omega}_{ri}$ and $\boldsymbol{\omega}_{si}$ are respectively relative rolling and twisting angular velocities from particle $j$ to $i$; $r^*$ is the reduced radius. Using radii $r_i$ and $r_j$ of particle $i$ and $j$, $r^*$ is defined as $\frac{1}{r^*} = \frac{1}{r_i} + \frac{1}{r_j}$. The coefficient values in Eq. (4) are taken directly from the values determined by Jiang et al. (2015).

### 2.2. Simulation Setup

This study investigates the tip angle dependence of resistive force when a cone vertically penetrates a granular layer at a constant velocity. The simulation setup is shown in Fig. 1 (a). The layer consists of granular particles with an average diameter of 2 mm, and the dimensions are 300 mm in width and depth and 180 mm in height, respectively. The edge of width ($x$-axis) and depth ($y$-axis) directions are periodic boundaries. Thereby, particles moving to coordinates below 0 mm or above 300 mm appear from the opposite sides. In the height ($z$-axis) direction, the fixed flat floor is set at the bottom of layer ($z = 0$ mm) to prevent particles from falling. This study set this fixed floor to the same parameters as the particles to minimize the wall effect. In contrast, the surface of a granular layer ($z = 180$ mm) is not set any boundary condition, thus, it allows particles move freely near the surface of a layer.

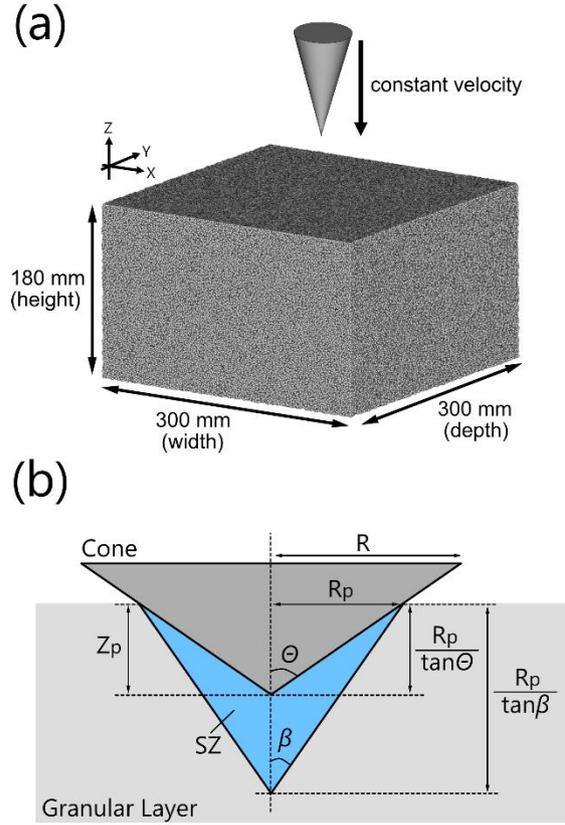

Fig. 1. (a) Schematic of simulation setup. (b) Schematic in cross-section view of the cone penetrating to granular layer.

Before the cone penetration simulation is carried out, granular layer is firstly prepared. The granular layers are adjusted to $\psi = 0.60$ for all conditions. After preparation, the cone positioned with the center of the surface penetrates to the layer at constant velocity 50 mm/s. The cone shape is defined by the radius $R$ and the tip angle $\Theta$, as shown in Fig. 1 (b). In this study, five different cones are used to investigate the tip angle effect on penetration resistive forces. The specific values ($\Theta$, $R$) of five cones are (15 deg, 28.6 mm), (30 deg, 28.6 mm), (45 deg, 42.9 mm), (60 deg, 42.9 mm) and (75 deg, 42.9 mm), respectively. As shown in the Fig. 1 (b), $z_p$ is defined as the depth to the cone-tip from the initial free surface level of granular layer. Cone penetrations are performed until the cone is completely buried in the layer.

In the simulation, this study uses four different parameter sets varying $\mu_{gg}$ and $\chi$, and each condition is shown in Table 1. The sand piles created using these parameter sets are shown in Fig. 2. Calculating $\phi$ from the profile of sand pile using the least squares method, $\phi$ are obtained as 17.3 deg for Type A, 25.0 deg for Type B, 35.6 deg for Type C, and 38.6 deg for Type D, respectively.

*Table 1. DEM parameter sets*

| Material Type | A | B | C | D |
|---|---|---|---|---|
| Young's modulus $E$ [Pa] | $1.0 \times 10^9$ | $1.0 \times 10^9$ | $1.0 \times 10^9$ | $1.0 \times 10^9$ |
| Poisson ratio $\nu$ [-] | 0.25 | 0.25 | 0.25 | 0.25 |
| coefficient of restitution $e$ [-] | 0.9 | 0.9 | 0.9 | 0.9 |
| particle diameter $d_1, d_2, d_3$ [mm] | 1.7, 2.0, 2.3 | 1.7, 2.0, 2.3 | 1.7, 2.0, 2.3 | 1.7, 2.0, 2.3 |
| particle mixing ratio $d_1:d_2:d_3$ [-] | 1 : 2 : 1 | 1 : 2 : 1 | 1 : 2 : 1 | 1 : 2 : 1 |
| particle density $\rho_g$ [kg/m³] | 2500 | 2500 | 2500 | 2500 |
| object density $\rho_o$ [kg/m³] | 2700 | 2700 | 2700 | 2700 |
| friction coefficient $\mu_{gg}$ [-] (particle - particle) | 0.1 | 0.2 | 0.8 | 1.0 |
| friction coefficient $\mu_{og}$ [-] (object - particle) | 0.3 | 0.3 | 0.3 | 0.3 |
| shape parameter $\chi$ [-] | 0.05 | 0.2 | 0.6 | 1.0 |
| timestep $\Delta t$ [s] | $4.0 \times 10^{-6}$ | $4.0 \times 10^{-6}$ | $4.0 \times 10^{-6}$ | $4.0 \times 10^{-6}$ |

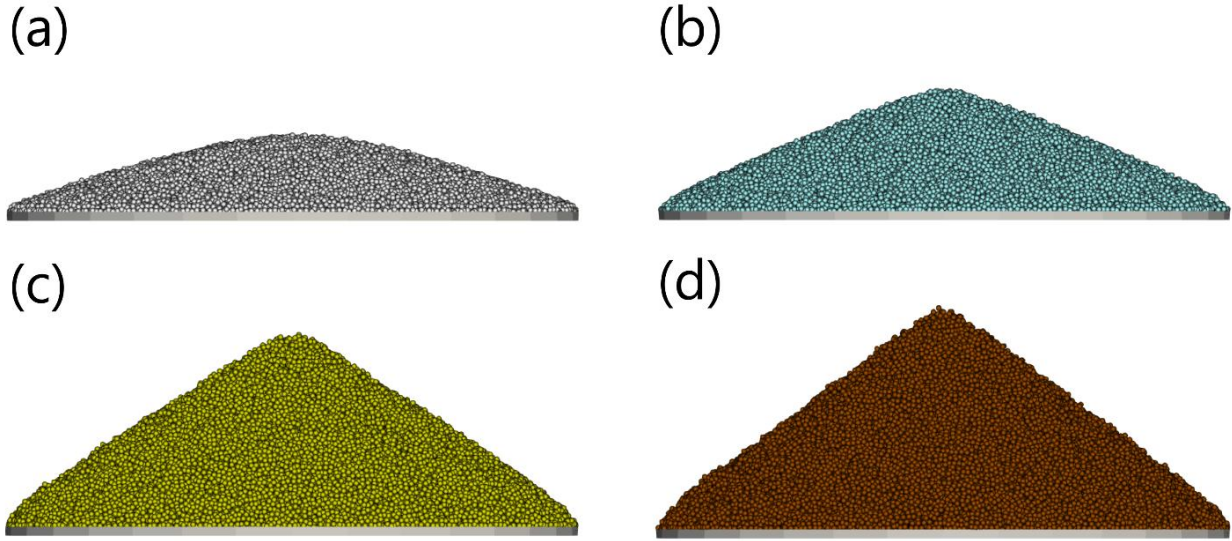

*Fig. 2. Sandpiles for (a) Type A, (b) Type B, (c) Type C, and (d) Type D.*

## 3. Results and Discussions

The results of the simulated resistive forces $F_{z,DEM}$ are firstly compared with *Eq. (1)*. Here, $V_p$ at the depth $z_p$ in *Eq. (1)* is calculated as follows:

$$V_p(z_p) = \frac{1}{3}\pi z_p^3 \tan^2\Theta. \tag{5}$$

Moreover, $K_\phi$ is calculated from *Eq. (1)* using $\phi$, $K_\phi = 7.3$ for Type A, $K_\phi = 20.0$ for Type B, $K_\phi = 103.7$ for Type C and $K_\phi = 177.8$ for Type D, respectively. *Figure 3* shows $F_{z,DEM}$ with colored lines and the corresponding estimation by MALT with dashed red lines. In the simulation results, the color shades represent the difference of $\Theta$. The horizontal axis is $z_p$ and the vertical axis is the resistive force in log-log scale in *Fig. 3*. In *Fig. 3*, $F_{z,DEM}$ is proportional to the cube of $z_p$, and qualitatively consistent with the MALT. However, for large $\Theta$, $F_{z,DEM}$ quantitatively deviates from MALT.

To identify and evaluate deviations between $F_{z,DEM}$ and MALT regarding $\Theta$, this study defines the ratio $D_{MALT}$ as follows:

$$D_{MALT} = \frac{F_{z,DEM}}{K_\phi \rho_g \psi g V_p}. \tag{6}$$

The values of $D_{MALT}$ are computed using the last 80 % of $F_{z,DEM}$ for each $\Theta$. The computed $D_{MALT}$ is shown by hollow markers in *Fig. 4*. Colors and markers indicate Types presented in *Table 1*. *Figure 4* indicates that the hollow markers increase when $\Theta$ is larger than around 45 deg. The increase magnitudes depend on $\phi$, with larger $\phi$ having a tendency towards larger values of $D_{MALT}$ for blunt cones. If there is no effect of cone-tip shape to the resistive forces, $D_{MALT}$ should be almost constant regardless of $\Theta$. Therefore, this result reveals penetration resistive forces are affected by the cone-tip shapes.

The cone-tip-angle-dependent changes in penetration resistive forces are expected to be due to a stagnant zone (SZ) formed in front of the intruder penetrating to the granular layer. Here, SZ is defined as a region in which granular particles move as a rigid body in front of the intruder. In fact, Aguilar and Goldman (2016) have confirmed the formation of SZ in front of cylinder penetrating to the granular layer through experimental analysis of particle motion. Moreover, for the flat-bottom cylinder, Aguilar and Goldman (2016); Kang et al (2018) have confirmed that a sudden increase in the resistive forces occurs in the shallow penetration depth (about 0.1 times the intruder's diameter) where the SZ develops in front of cylinder.

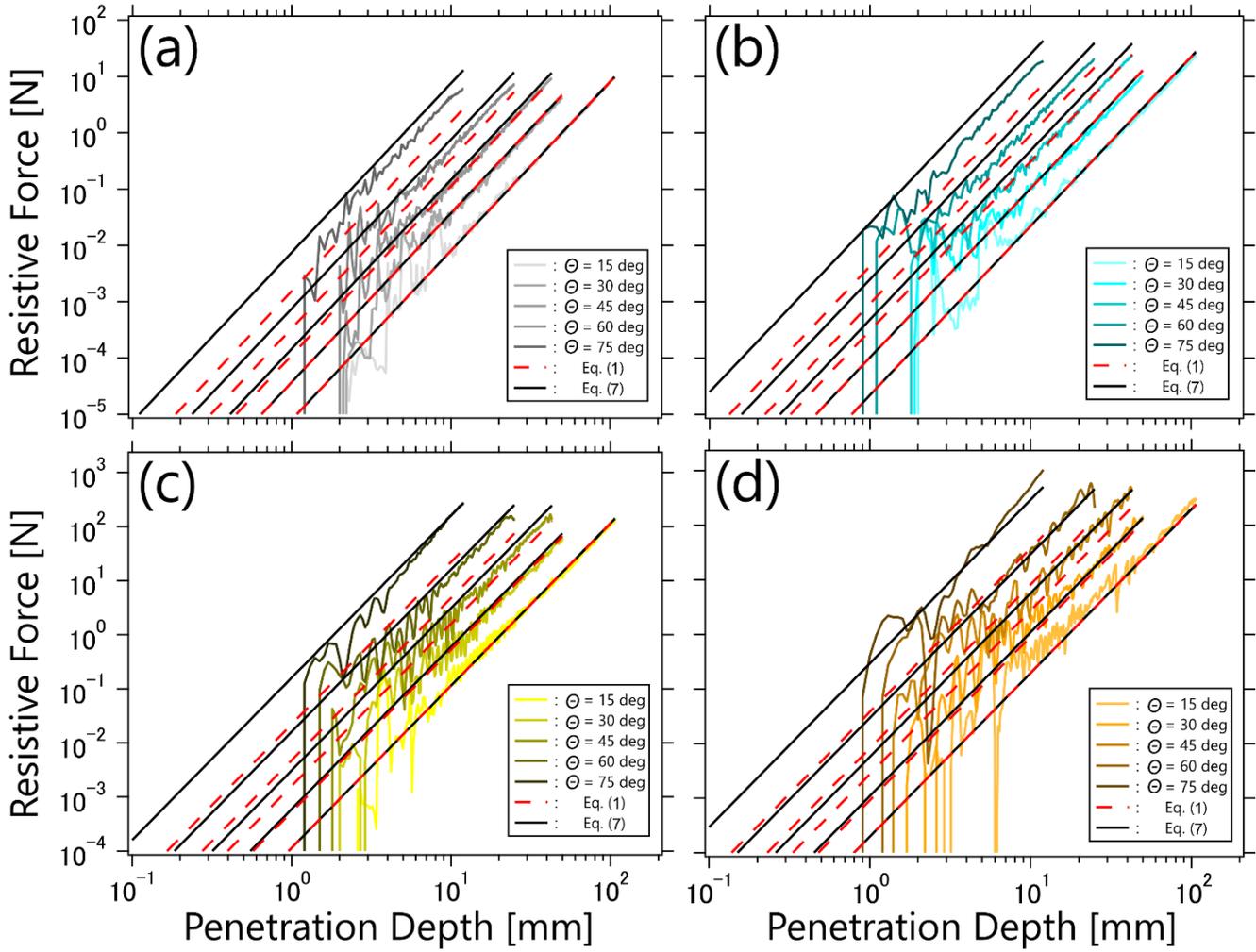

*Fig. 3. Relationship between resistive force and penetration depth for (a) Type A, (b) Type B, (c) Type C, and (d) Type D.*

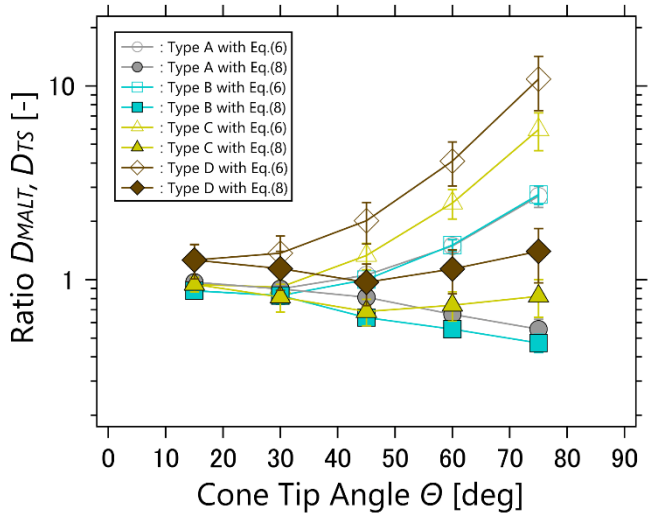

*Fig. 4. Variation of $D_{MALT}$ and $D_{TS}$ with respect to $\Theta$.*

This study investigates the particle velocity fields around a cone to clarify the formation of SZ and the effect of the tip shape on resistive forces. The particle velocity fields for Type B in the final states of cone penetration with different tip angles are shown in *Fig. 5*. Here, the colors of the particle, red and blue, indicate positive and negative particle velocities in *z*-direction, respectively. The tip angle of SZ, $\beta$ is defined as $\beta = \frac{\pi}{4} - \frac{\phi}{2}$, as in Kang et al (2018), and $\beta = 32.5$ deg for Type B. *Figure 5* shows that the distribution of the particle velocities in front of the cone varies depending on tip shape. In the case of cone tips sharper than SZ ($\Theta$ = 15 and 30 deg), most of the particles in front of the cone are not moving in the penetration direction. On the other hand, in the case of a blunt cone ($\Theta$ = 45, 60 and 75 deg), particles in front of the cone, especially within the SZ, are moving in the penetration direction. Thus, SZ develops in front of the cone when blunt cones penetrate.

From the above result, this study explains the discrepancy between $F_{z,DEM}$ and MALT by considering SZ as a virtual mass. Indeed, Aguilar and Goldman (2016) have explained the large resistive force on the penetrating cylinder assuming SZ as a virtual mass. Specifically, for a blunter cone than SZ ($\Theta > \beta$), this study considers the situation shown as *Fig. 1 (b)* in which SZ develops in front of the blunt cone. The penetration depths of cone and SZ are respectively $\frac{R_p}{\tan\theta}$ and $\frac{R_p}{\tan\beta}$. Accordingly, the penetration volume of blunt cone assuming a virtual mass is the product of *Eq. (5)* and $\frac{\tan\theta}{\tan\beta}$.

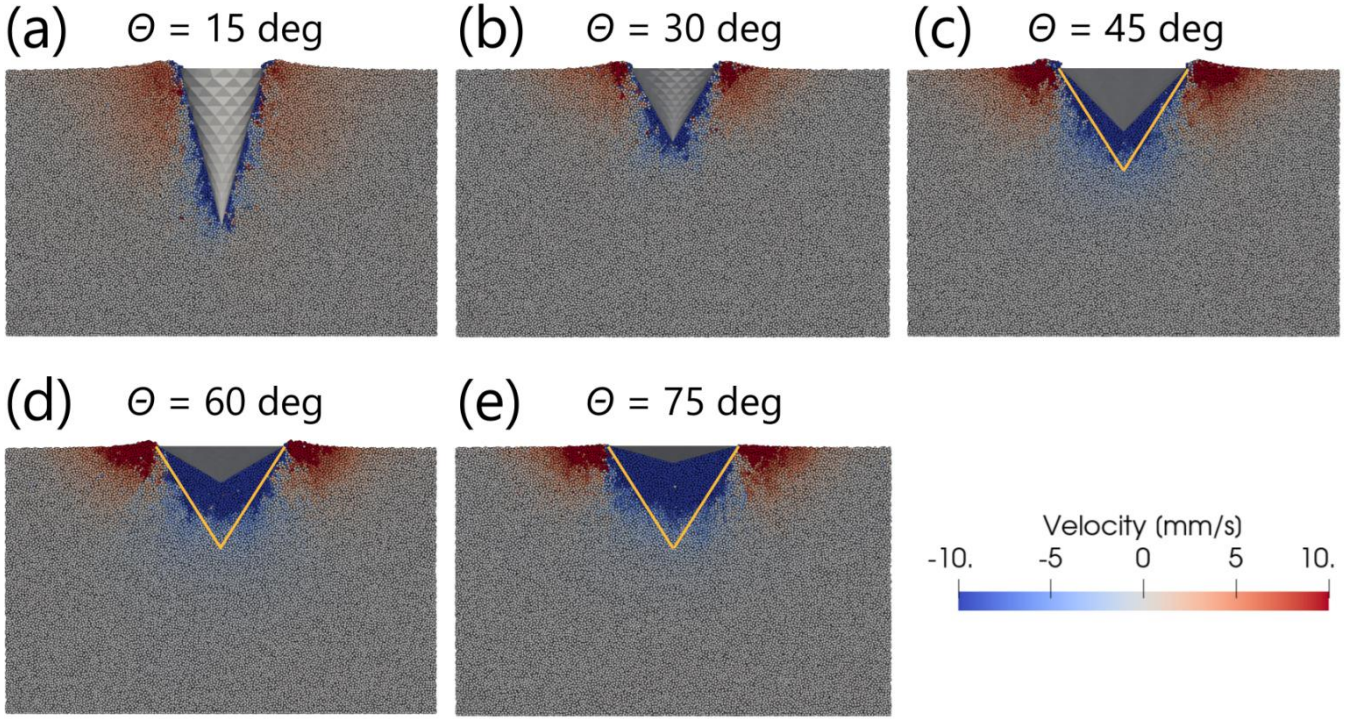

*Fig. 5. Particle velocity field in z-direction on cross-section view for Type B.*

Thus, *Eq. (1)* is modified by considering a virtual mass as the coefficient $f(\Theta)$, and it is as follows:

$$\boldsymbol{F}_z(z_p) = f(\Theta)\, K_\phi \rho_g \psi \boldsymbol{g} V_p(z_p) \tag{7}$$

with

$$f(\Theta) = \begin{cases} 1 & : \text{if } \Theta \leq \beta \\ \frac{\tan\Theta}{\tan\beta} & : \text{if } \Theta > \beta \end{cases}.$$

The estimation results of *Eq. (7)* are shown with solid black lines in *Fig. 3*. Comparing the black solid and red dashed lines in *Fig. 3*, the black lines are closer to $\boldsymbol{F}_{z,DEM}$ even for blunt cone. To identify and evaluate deviations between $\boldsymbol{F}_{z,DEM}$ and *Eq. (7)*, this study further defines the ratio $D_{TS}$ as follows:

$$D_{TS} = \frac{\boldsymbol{F}_{z,DEM}}{f(\Theta) K_\phi \rho_g \psi \boldsymbol{g} V_p}. \tag{8}$$

$D_{TS}$ is also shown by solid markers in *Fig. 4*. Colors and markers indicate Types of the penetrated granular layer.

From *Fig. 4,* while $D_{MALT}$ increases with $\Theta$, $D_{TS}$ decreases with $\Theta$ or is almost 1 regardless of $\Theta$. The $D_{TS}$ behavior slightly depends on the material types. The $D_{TS}$, is almost 1 for Type C and D regardless of $\Theta$, while it decreases with $\Theta$ for Type A and B. The lower limit of the estimated $D_{TS}$ is around 0.5, which means that the difference between *Eq. (7)* and $\boldsymbol{F}_{z,DEM}$ is about twice as large. By contrast, the largest value of $D_{MALT}$ is about 10 for blunt cones, which indicates that the ratio between *Eq. (1)* and $\boldsymbol{F}_{z,DEM}$ can be about 10 times. Therefore, the proposed model can more precisely estimate penetration resistive force to dry granular layers in most of the cases than MALT, although it may slightly overestimate the value depending on the cone tip shapes and material types.

Incidentally, the assumption for increase of penetration resistive force due to the virtual mass by SZ needs further verification. This study uses this assumption since the analysis of particle velocity field for Type B shows that particle moving zones in front of the cone match the region of the predicted SZ. However, the proposed model, which assumes an increase of penetration volume due to a virtual mass, overestimates the resistive forces on blunt cones in Type B compared to $\boldsymbol{F}_{z,DEM}$. This result means that the virtual mass produced by SZ does not fully explain $\boldsymbol{F}_{z,DEM}$. To discuss this point, more detailed studies, including an analysis of the failure mechanism of the granular layer, will be needed.

## 4. Conclusions and future work

This study investigated the tip angle dependence for resistive force into dry granular materials for shallow cone penetration at a constant velocity. The resistive force was calculated through DEM simulation and compared with MALT model (*Eq. (1)*). When the cone had shaper tip than the SZ, MALT can estimate the restive forces. On the other hand, when the cone had blunter tip than the SZ, MALT underestimated the resistive forces. Through the analysis of the particle velocity field around the penetrated cone, it is revealed that the increase in the resistive forces for blunt cone was caused by a virtual mass due to SZ. From this result, this study proposed a modified model considering the coefficient $f(\Theta)$ in the MALT that considered the growing volume of the SZ. In addition, comparing with the results of DEM simulation, the proposed model was validated. As a result, the proposed model can estimate the resistive force more precisely than MALT.

In this study, the resistive forces to simple intruders against a dry granular layer was investigated. However, in practical applications, it is necessary to consider more complex situations such as the intruder shape other than cones, the grain size distribution and cohesion in the ground, as well as the effects of their spatial heterogeneity. Therefore, future study should extend the proposed models for various effects and validate their effectiveness through experiments.

## 5. Nomenclature

| | | |
|---|---|---|
| $F_n^j, F_t^j$ | Normal and tangential forces between particle $i$ and $j$ | [kg m/s$^2$] |
| $g$ | Gravitational acceleration | [m/s$^2$] |
| $I_i$ | Moment of inertia of particle $i$ | [kg/m$^2$] |
| $k_n, k_t$ | Normal and tangential spring constants | [kg/s$^2$] |
| $K_\phi$ | Coefficients of friction-derived forces | [-] |
| $m_i$ | Mass of particle $i$ | [kg] |
| $M_r^j, M_s^j$ | Rolling and twisting moments from particle $j$ to $i$ | [kg m$^2$/s$^2$] |
| $R$ | Radius of cone | [m] |
| $r^*$ | Reduced particle's radius | [m] |
| $r_i^j$ | Vector from the center of particle $i$ to the contact point with particle $j$ | [m] |
| $R_p$ | Penetrated radius of cone | [m] |
| $S_n, S_t$ | Coefficients to caluculate $\gamma_n$, and $\gamma_t$ | [-] |
| $v_i$ | Velocity of particle $i$ | [m/s] |
| $V_p$ | Penetration volume of a intruder | [m$^3$] |
| $v_{nij}, v_{tij}$ | Normal and tangential relative velocities from particle $i$ to $j$ | [m/s] |
| $z_p$ | Penetration depth | [mm] |
| $\beta$ | Tip angle of stagnant zone | [-] |
| $\delta_{\theta ri}, \delta_{\theta si}$ | Total relative rolling and twisting angular displacements from particle $j$ to $i$ | [-] |
| $\delta_{nij}, \delta_{tij}$ | Normal and tangential displacements between particle $i$ and $j$ | [m] |
| $\gamma_n, \gamma_t$ | Normal and tangential viscous dampings | [kg/s] |
| $\mu$ | Sliding friction coefficient | [-] |
| $\omega_i$ | Angular velocity of particle $i$ | [1/s] |
| $\omega_{ri}, \omega_{si}$ | Relative rolling and twisting angular velocities from particle $j$ to $i$ | [1/s] |
| $\phi$ | Internal friction angle | [deg] |
| $\Theta$ | Tip angle of cone | [deg] |
| $\xi$ | Coefficient to calculate $\gamma_n$ and $\gamma_t$ | [-] |

## 6. Acknowledgements

N. Iikawa would like to acknowledge the financial support from Komatsu Ltd.

## 7. Declaration of competing interest

The authors declare that they have no known competing financial interests or personal relationships that could have appeared to influence the work reported in this paper.